\documentclass[aps,prb,reprint,10pt,superscriptaddress,amssymb,amsfonts,longbibliography]{revtex4-1}

\usepackage{amsmath}
\usepackage{graphicx}
\usepackage{dcolumn}
\usepackage{bbm}
\usepackage{subfigure}
\usepackage{amssymb}
\usepackage[colorlinks,bookmarks=false,citecolor=blue,linkcolor=red,urlcolor=blue]{hyperref}
\usepackage{wasysym}
\usepackage{MnSymbol}
\usepackage{color}
\usepackage{braket}

\stepcounter{secnumdepth}
\stepcounter{tocdepth}

\begin{document}

\title{Exactly Solvable Model for Two Dimensional Topological Superconductor}

\author{Zitao \surname{Wang}}
\affiliation{Department of Physics and Institute for Quantum Information and Matter, California Institute of Technology, Pasadena, California 91125, USA}
\author{Shang-Qiang \surname{Ning}}
\affiliation{Institute for Advanced Study, Tsinghua University, Beijing 100084, China}

\author{Xie \surname{Chen}}
\affiliation{Department of Physics and Institute for Quantum Information and Matter, California Institute of Technology, Pasadena, California 91125, USA}

\begin{abstract}
In this paper, we present an exactly solvable model for two dimensional topological superconductor with helical Majorana edge modes protected by time reversal symmetry. Our construction is based on the idea of decorated domain walls and makes use of the Kasteleyn orientation on a two dimensional lattice, which was used for the construction of the symmetry protected fermion phase with $Z_2$ symmetry in Ref. \onlinecite{Tarantino2016,Ware2016}. By decorating the time reversal domain walls with spinful Majorana chains, we are able to construct a commuting projector Hamiltonian with zero correlation length ground state wave function that realizes a strongly interacting version of the two dimensional topological superconductor. From our construction, it can be seen that the $T^2=-1$ transformation rule for the fermions is crucial for the existence of such a nontrivial phase; with $T^2=1$, our construction does not work.
\end{abstract}

\maketitle

\textit{Introduction} -- The discovery of topological insulators and superconductors\cite{Kane2005,Fu2007,Moore2007,Roy2009,Roy2008arXiv,Qi2009} demonstrates that a fermionic system can exhibit nontrivial topological properties if the fermions occupy a band structure with nontrivial topology. In particular, it was realized that the topological insulators and superconductors host gapless edge modes around a gapped bulk, which cannot be removed unless certain symmetry is explicitly or spontaneously broken. Moreover, the topological nature of the phases is also manifested at symmetry defects on the boundary of the system. For example, in a 2D topological superconductor, a time reversal domain wall on the 1D boundary hosts a Majorana zero mode and in a 3D topological superconductor, a time reversal domain wall on the 2D boundary hosts a chiral Majorana mode. A complete classification of topological insulators and superconductors in free fermion systems was given in Ref.\onlinecite{Schnyder2008,Kitaev2009}.

Such `Symmetry Protected Topological (SPT)' order was generalized to boson systems as well, although in a very different setting. It was discovered that while without interaction bosons systems cannot have symmetry protected gapless modes around a gapped bulk, with strong interaction, a large variety of SPT orders is possible. A whole class of exactly solvable models with commuting projector Hamiltonian and zero correlation length ground state wave function were constructed to realize such bosonic SPT order\cite{Chen2012, Chen2013}. 

Can topological insulators and superconductors discovered in the free fermion setup be realized with exactly solvable models as well? This question is interesting not only out of pure theoretical curiosity; it is also crucial for formulating a general framework for both fermionic and bosonic SPT phases which may lead to the discovery of new phases and a complete classification. Moreover, it can be useful in answering questions regarding many-body localization in such phases when strong disorder is present\cite{Potter2015arXiv}. In this paper, we focus on the case of 2D topological superconductor.

If an exactly solvable model is possible, it necessarily involves interactions as the free fermion ground states always have a nonzero correlation length due to the nontrivial topology of the band structure. Ref. \onlinecite{Gu2014,Wang2017arXiv} gave the exactly solvable model realization of a large class of fermionic SPT phases which are protected by symmetry of the form $G_b \times Z_2^f$, where $G_b$ denotes symmetry transformation on some bosonic degrees of freedom in the system and $Z_2^f$ is the fermion parity part of the symmetry. The symmetry protecting the topological superconductor falls out of this class. In the topological superconductor, time reversal symmetry acts as $T^2=P_f$, where $P_f$ is the fermion parity operator generating the $Z_2^f$ symmetry group. Therefore, the total symmetry group is $Z_4$, with the odd group elements being anti-unitary.

\begin{figure}[tbh]
\includegraphics[width=.45\textwidth]{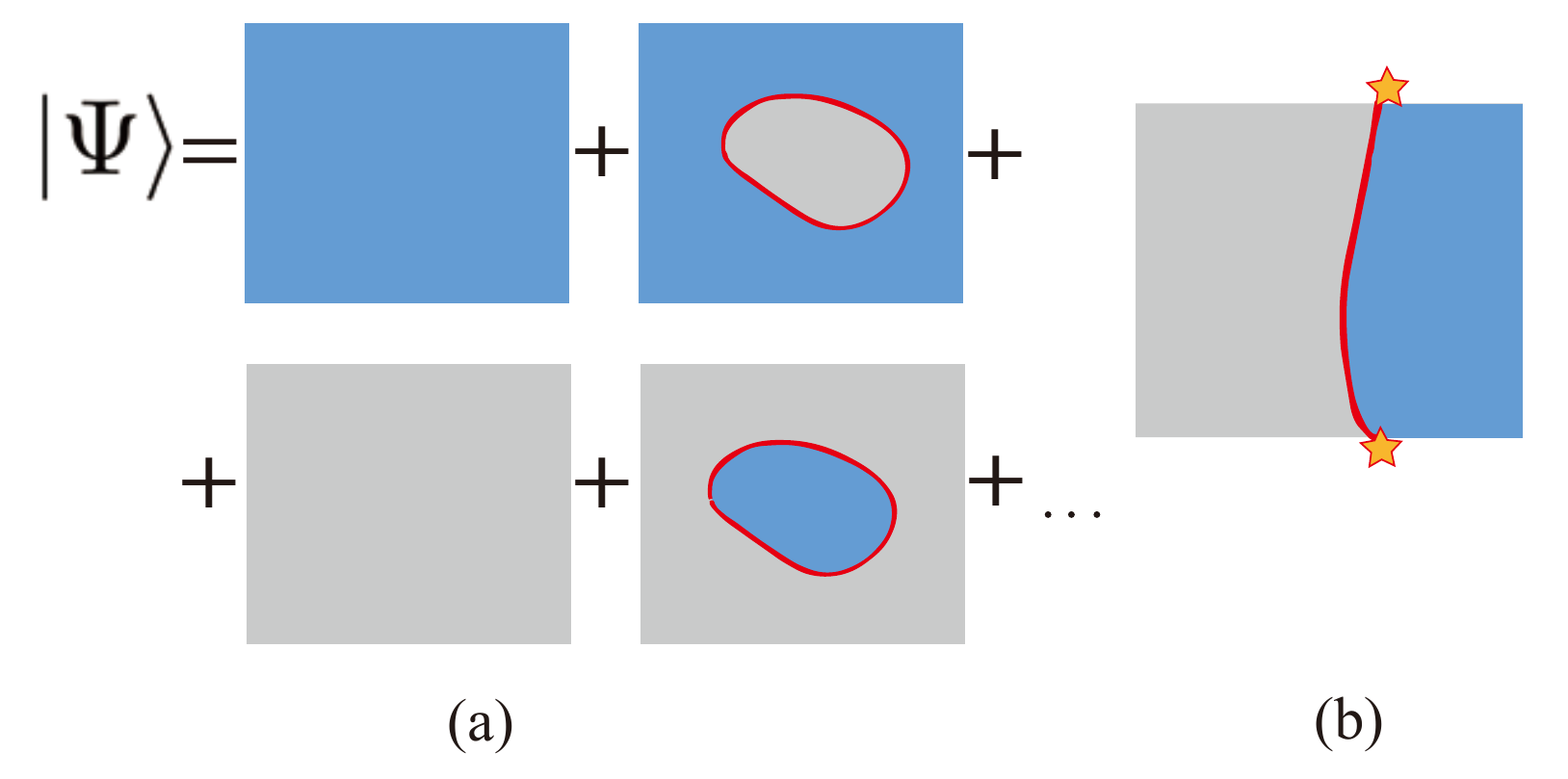}
\caption{The decorated domain wall approach. (a) Ground state is a superposition of all symmetry breaking domain configurations (blue and grey patches) with domain walls decorated with SPT states of one lower dimension (red curves). (b) The end point of the domain wall on the boundary (star) hosts nontrivial edge states of the lower dimensional SPT.
}
\label{fig:DDW}
\end{figure}

The decorated domain wall construction provides a different approach for constructing exactly solvable models for SPT phases.\cite{Chen2014}. In this approach, the ground state wave function is written as a superposition of all possible symmetry breaking configurations with the symmetry breaking domain walls being decorated with SPT states of one lower dimension, as shown in Fig.\ref{fig:DDW} (a). The superposition guarantees that the total wave function is symmetric. Moreover, when symmetry is broken into opposite domains, the domain wall carries the lower dimensional SPT state. When the domain wall ends on the boundary of the system, the end point hence hosts the edge state of the lower dimensional SPT state, reflecting the nontrivial nature of the original SPT order, as shown in Fig.\ref{fig:DDW} (b). 

In a topological superconductor with helical Majorana edge mode described by $H_{\text{edge}} = \sum_{k_y \geq 0} v_Fk_y \left(\psi^{\dagger}_{k_y\uparrow}\psi_{k_y\uparrow} - \psi^{\dagger}_{k_y\downarrow}\psi_{k_y\downarrow}\right)$, a mass term of the form $\delta H = m\sum_{k_y\geq 0}\left(\psi^{\dagger}_{k_y\uparrow}\psi_{k_y\downarrow}+\psi^{\dagger}_{k_y\downarrow}\psi_{k_y\uparrow}\right)$
can gap out the edge mode while breaking time reversal symmetry. On the domain wall between $\delta H$ and $-\delta H$, there is an isolated Majorana mode. Therefore, if the topological superconductor can be written in the decorated domain wall way, we should decorate the time reversal domain walls with Majorana chains. 

Decorating symmetry domain walls with Majorana chains has proven to be more difficult than with bosonic chains. A breakthrough was made recently in Ref.\onlinecite{Tarantino2016,Ware2016} where a fermionic SPT phase with $Z_2 \times Z_2^f$ symmetry was realized by decorating the $Z_2$ domain walls with 1D Majorana chains. Although the protecting symmetry is still of the form $G_b\times Z_2^f$, this particular phase cannot be realized using the method of Ref.\onlinecite{Gu2014}. It was realized that the incorporation of a Kasteleyn orientation on the two dimensional lattice, which corresponds to a discrete version of spin structure in 2D, is crucial for a consistent decoration.

Using the Kasteleyn orientation, we present a decorated domain wall construction of the 2D topological superconductor in this paper. Our construction is different from that of the $Z_2 \times Z_2^f$ SPT phase in an important way. In the case of $Z_2 \times Z_2^f$, the Majorana chain used for decoration does not transform under the $Z_2$ part of the symmetry, which acts only on the symmetry domains. In the case of topological superconductor, time reversal acts both on the symmetry domains and on the Majorana chains to be decorated onto the symmetry domain walls. In fact, the way the Majorana chains transform under time reversal is crucial for the construction as we know that topological superconductivity only exists for $T^2=-1$ fermions but not the $T^2=+1$ ones. Indeed, after we present carefully how a zero correlation length wave function and a commuting projector Hamiltonian can be constructed for $T^2=-1$ fermions, we will be able to see why a similar construction fails for the $T^2=+1$ ones.

\begin{figure}
\centering
\includegraphics[width = 0.9\linewidth]{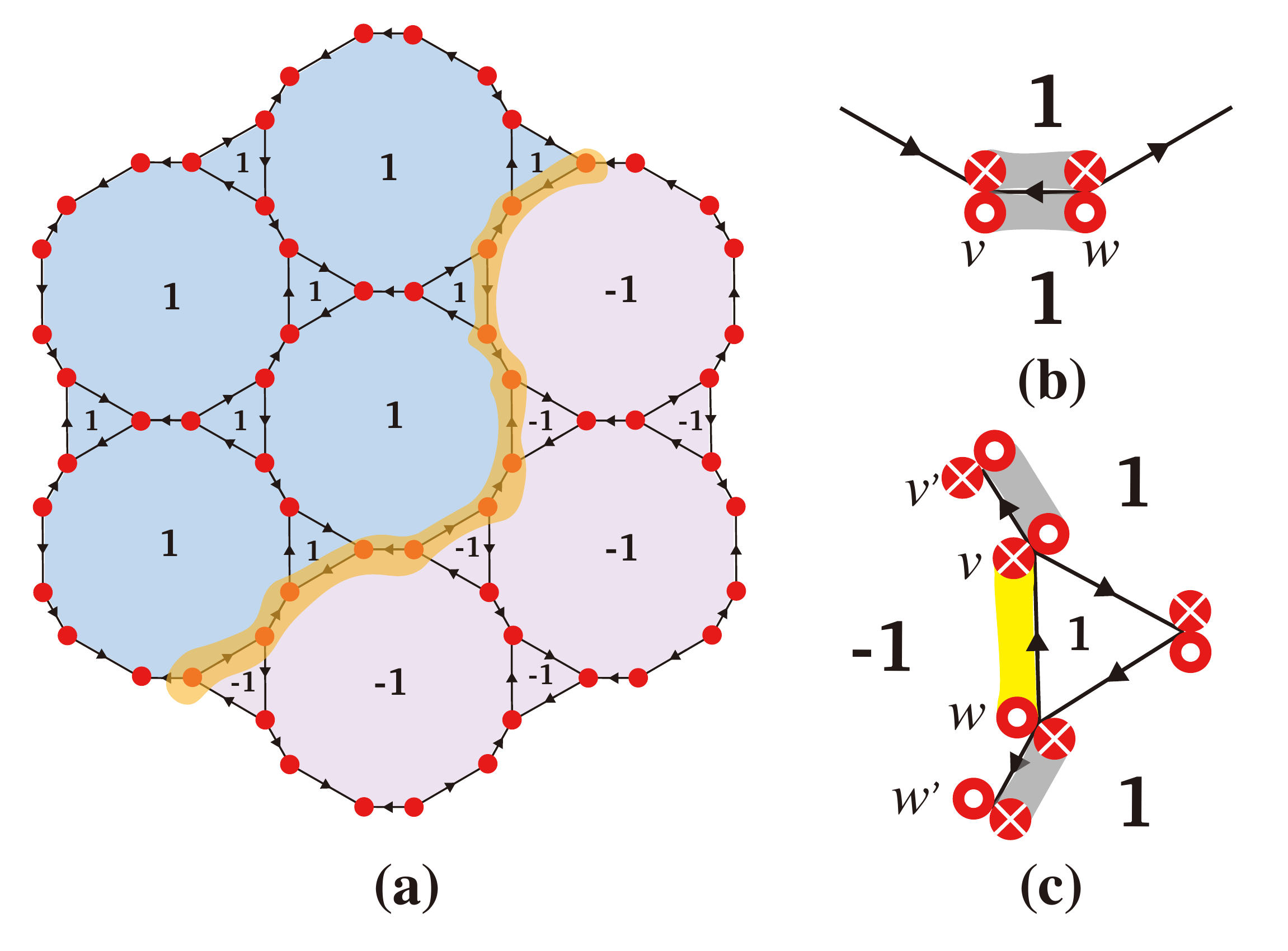}
\caption{(a) illustrates the lattice structure and degrees of freedom in our model. Here $1$ and $-1$ denote the eigenstates of $\tau^z_p$ with eigenvalues $1$ and $-1$, respectively. The blue bonds indicate the time-reversal domain wall. The solid red circles denote the Majorana modes $\gamma_{v}^{\sigma}$ ($\sigma = \uparrow, \downarrow$). The arrow at each bond denotes the Kasteleyn orientation of the bond. (b) (resp. (c)) is a detailed illustration of the coupling of Majorana modes away from (resp. on) the domain wall. The dots and crosses on the solid red circles indicate the up ($\uparrow$) and down ($\downarrow$) spins of the Majorana modes, respectively. The yellow (resp. grey) bond denotes the coupling of Majorana modes that share a long (resp. short) bond.}
\label{dof}
\end{figure}

\textit{Wave-function} -- 
Consider the planar trivalent lattice in Fig.\ref{dof} together with a Kasteleyn orientation, i.e., orientation of the bonds of the lattice for which any plaquette has an odd number of clockwise-oriented bonds. There are two types of faces in the lattice: the $12$-sided faces, which we will refer to as plaquettes, and the triangular faces, which we will refer to as triangles. Let $t(v)$ and $t(w)$ be the triangles that contain the vertices $v$ and $w$, respectively. The bonds of the lattice also come in two types: The `short' bonds which connect different triangles  ($t(v) \neq t(w)$), and the `long' bonds that are in the same triangle  ($t(v) = t(w)$).

The Hilbert space of our model consists of a bosonic spin-$1/2$ located on each plaquette $p$, acted on by the Pauli operators $\tau_p^x$, $\tau_p^y$, $\tau_p^z$, and a pair of complex fermions located on each short bond $l$, created and annihilated by operators $c_l^{\sigma \dagger}$ and $c_l^{\sigma}$ ($\sigma = \uparrow, \downarrow$), respectively. Let $l = \langle \overrightarrow{vv^{\prime}} \rangle$ be oriented from vertex $v$ to vertex $v^{\prime}$. Each complex fermion on $l$ can be represented by a pair of Majorana modes
\begin{align}
\gamma^{\sigma}_{v} &= c^{\sigma \dagger}_{l} + c^{\sigma}_{l}, \nonumber \\
\gamma^{\sigma}_{v^{\prime}} &= i(c^{\sigma \dagger}_{l}-c^{\sigma}_{l}),
\label{Majorana}
\end{align}
located at $v$ and $v^{\prime}$, respectively. We can also define a fictitious spin-$1/2$ degree of freedom $\tau_t$ on each triangle following the majority rule: The value of $\tau_t$ is set to $1$ if the majority of the three plaquettes bordering $t$ have $\tau_p^z = 1$, and is set to $-1$ otherwise.

Our system has a time reversal symmetry $T$, which acts on both the plaquette spins and the complex fermions. In the eigenbasis of $\tau_p^z$, $T$ maps between the two eigenstates of $\tau^z_p$:
\begin{align}
T: \ket{1} \rightarrow \ket{-1}, \ \ \ \ket{-1} \rightarrow \ket{1},
\end{align}
together with the complex conjugation operation in this basis. The fictitious spins on the triangles will also be flipped due to the majority rule. Since any fixed plaquette spin configuration in the $\tau^z$ basis breaks time reversal symmetry, we will refer to a domain of plaquette spins in the same $\tau^z$ basis state as a time reversal domain. Furthermore, $c_l^{\sigma}$ transforms as a Kramers doublet under $T$:
\begin{align}
T: c_l^{\uparrow} \rightarrow c_l^{\downarrow}, \ \ \ c_l^{\downarrow} \rightarrow -c_l^{\uparrow}.
\end{align}
Written in terms of the Majorana modes, we have:
\begin{align}
T: 
&\begin{cases}
\gamma_v^{\uparrow} \rightarrow \gamma_v^{\downarrow}\\
\gamma_v^{\downarrow} \rightarrow -\gamma_v^{\uparrow},
\end{cases} 
\begin{cases}
\gamma_{v^{\prime}}^{\uparrow} \rightarrow -\gamma_{v^{\prime}}^{\downarrow}\\
\gamma_{v^{\prime}}^{\downarrow} \rightarrow \gamma_{v^{\prime}}^{\uparrow}.
\end{cases} 
\end{align}
where the Kasteleyn orientation points from $v$ to $v'$.

Now we describe in detail how we decorate the time reversal domain walls with Majorana chains. 
Away from the domain wall, we pair up Majorana modes that share a short bond $\langle \overrightarrow{vv^{\prime}} \rangle$
as $i\gamma_v^{\uparrow} \gamma_{v^{\prime}}^{\uparrow} + i\gamma_v^{\downarrow} \gamma_{v^{\prime}}^{\downarrow}$. On a domain wall, we pick out one Majorana mode $\gamma_v^{\sigma_v}$ from each vertex $v$ and pair them along the long bonds $\langle \overrightarrow{vw} \rangle$ as $i\gamma_v^{\sigma_v}\gamma_w^{\sigma_w}$ so that they form a Majorana chain.
The spin label $\sigma_v$ is determined as follows: We set up a local coordinate system on the short bond that contains $v$ by regarding the orientation of this bond as the $x$ axis in the 2D plane. The $y$ axis is then uniquely determined by the right-hand rule and the orientation of the 2D plane. If the $y$ axis points from the $\ket{-1}$ domain to the $\ket{1}$ domain, we set $\sigma_v =\ \uparrow$. Otherwise, we set $\sigma_v =\ \downarrow$. 
After the Majorana modes of the $\sigma_v$ species pair into Majorana chains,
we are left with exactly one unpaired Majorana mode on each vertex on the domain wall. The two unpaired Majorana modes that share a short bond $\langle \overrightarrow{vv'}\rangle$ will have the same spin $\bar{\sigma}_v$ which can be paired as 
$i\gamma_v^{\bar{\sigma}_v}\gamma_{v^{\prime}}^{\bar{\sigma}_{v^{\prime}}}$.
 This is the same kind of coupling as that away from the domain wall, but with only one species of Majorana modes. Fig.\ref{dof} (b) and (c) give a pictorial illustration of these coupling rules. 

The ground state wave function of a topological superconductor is then given by the superposition of all possible time-reversal domain configurations with domain walls decorated with Majorana chains. It satisfies the following properties: it's time reversal invariant, and every configuration in the superposition has the same fermion parity. The latter fact is ensured by the Kasteleyn orientation. The reason for this is very similar to that presented in Ref.\onlinecite{Tarantino2016,Ware2016} although here we have two species of fermion modes.

To see the time reversal invariance, we note that time reversal acts by flipping the plaquette spins, and transforms the Majorana modes in a way that conforms to the decoration rules introduced above. In particular, for Majorana modes not on a domain wall, they pair as $i\gamma^{\uparrow}_v\gamma^{\uparrow}_{v'}+i\gamma^{\downarrow}_v\gamma^{\downarrow}_{v'}$ on a short bond which is invariant under time reversal. 
For Majorana modes on a domain wall, the decoration rule says that the modes that form (do not form) Majorana chains flip their spin when the plaquette spins are flipped, which is consistent with the time reversal transformation action. Moreover, the pairing terms along the domain wall, whose signs are fixed by the Kasteleyn orientation, exactly map into each other under time reversal without any sign ambiguity. To see this, first notice that for the modes which do not form Majorana chains, the pairing maps from $i\gamma_v^{\sigma_v}\gamma_{v^{\prime}}^{\sigma_{v^{\prime}}}$ to $i\gamma_v^{\bar{\sigma}_v}\gamma_{v^{\prime}}^{\bar{\sigma}_{v^{\prime}}}$, which are both consistent with the Kasteleyn orientation. Secondly, for the modes that are involved in forming Majorana chains, one can check that the pairing term $i\gamma_v^{\sigma_v}\gamma_w^{\sigma_w}$ is mapped into $i\gamma_v^{\bar{\sigma}_v}\gamma_w^{\bar{\sigma}_w}$ which are both consistent with the Kasteleyn orientation.\footnote{The way $\gamma_v^{\sigma_v}$ transforms into $\gamma_v^{\bar{\sigma}_v}$ depends on the orientation of the short bond $ <vv'>$ and similarly for $w$. One can check that with all four orientation possibilities, this conclusion is always true.} 
Therefore, we can conclude that time reversal maps from one to another the decorated domain wall configurations in the superposition. The whole superposition is then time reversal invariant if the weight of the time reversal partner configurations are complex conjugate of each other. This will be demonstrated in detail in Appendix \ref{HTR}.



\textit{Hamiltonian} -- 
The Hamiltonian of our model can be written as 
\begin{equation}
H = H_{\text{decorate}} + H_{\text{tunnel}},
\end{equation}
where $H_{\text{decorate}}$ will be defined to realize the domain wall decoration described in the above section for each plaquette spin configuration, and $H_{\text{tunnel}}$ will be defined to tunnel between the different plaquette spin configurations.

More explicitly, let $D_{\langle \overrightarrow{vw} \rangle} = \frac{1}{2}\left(1-\tau^z_{f_{\overrightarrow{vw}}} \tau^z_{f_{\overrightarrow{vw}}^{\prime}}\right)$
be the operator which detects if the bond (either short or long) $\langle \overrightarrow{vw} \rangle$ is on a domain wall. $f_{\overrightarrow{vw}}$ and $f_{\overrightarrow{vw}}^{\prime}$ are the two faces that share the bond $\langle \overrightarrow{vw} \rangle$, which can be either plaquettes or triangles. 
The left-hand-side face of the bond $\langle\overrightarrow{vw}\rangle$ is denoted by $f_{\overrightarrow{vw}}$ while the right-hand-side one by $f'_{\overrightarrow{vw}}$.

 If $\langle \overrightarrow{vw} \rangle$ is a long bond, we denote by $ \langle \overline{vv'}\rangle  $ ($\langle \overline{ww'} \rangle$) the short bond that includes vertex $v$($w$).  The overline on top of $ \overline{vv'} $ means that if $v$ is oriented to $v'$,  $ \overline{vv'} =\overrightarrow{vv'} $, otherwise $ \overline{vv'} =\overrightarrow{v'v} $.
Therefore we can  define two operators $W_{vw}^{\pm}=\frac{1}{4}\left(1\pm\tau_{f_{\overline{vv'}} }^z \right)\left(1 \mp \tau^z_{f'_{{\overline{ww'}}}}\right)$
to determine which $\gamma_{v,w}^s$( $s=\uparrow, \downarrow$) to pair in the Majorana chain on the domain wall. More explicitly, if $W_{vw}^{+}=1$, $W_{vw}^{-}=0$,  then the pairing over the long bond $\langle \overrightarrow{vw} \rangle$ is $i\gamma_v^\uparrow \gamma_w^{\downarrow}$;     if $W_{vw}^{-}=1$, $W_{vw}^{+}=0$ it is $i\gamma_v^\downarrow \gamma_w^{\uparrow}$. If both are zero, $\langle \overrightarrow{vw} \rangle$ is not on a domain wall.

\begin{figure}
\centering
\includegraphics[width = \linewidth]{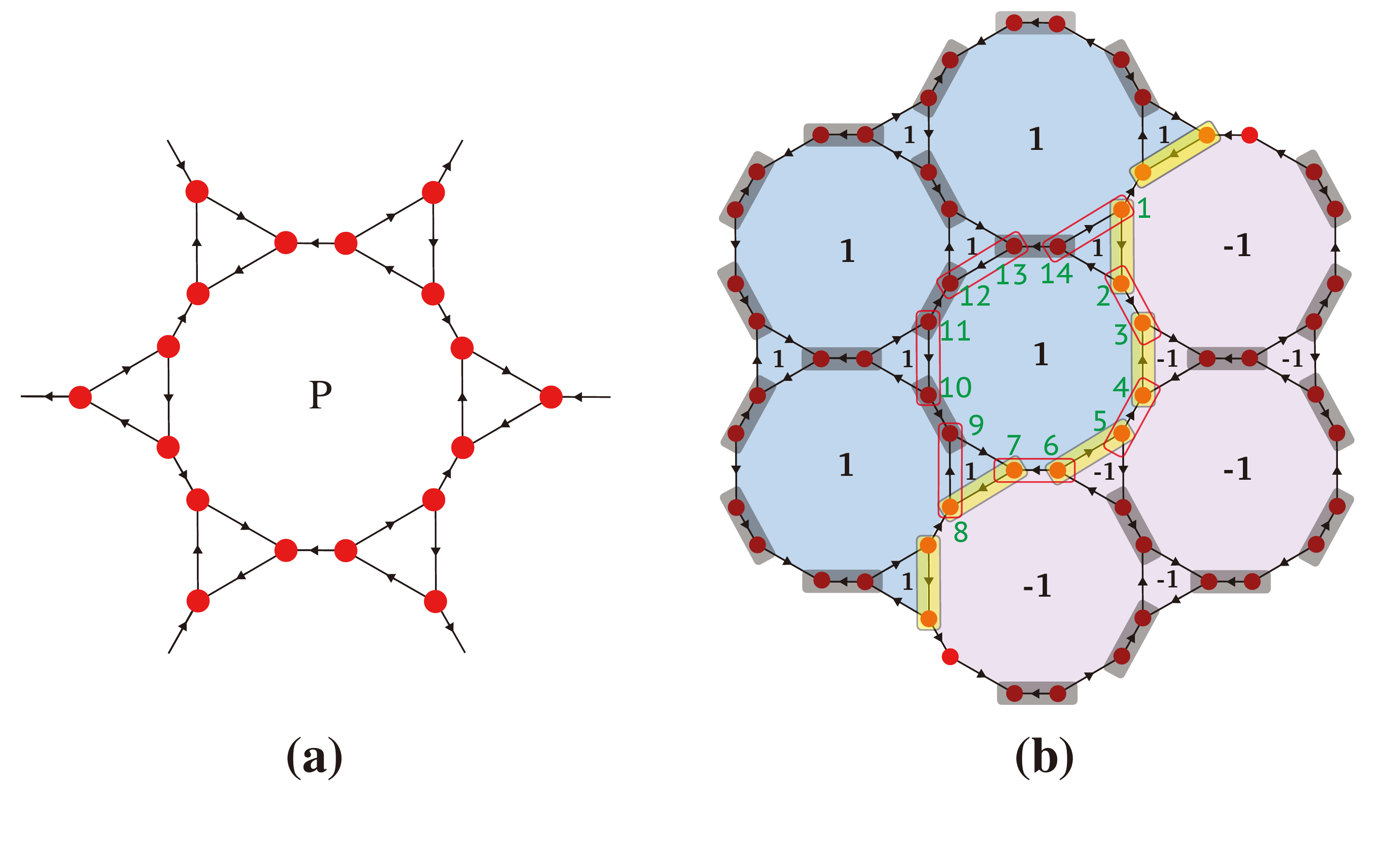}
\caption{(a) The $36$ Majorana modes denoted by the $18$ red dots in this figure are the Majorana modes surrounding the plaquette $p$, denoted by $\partial^{\prime} p$. (b) Majorana modes (labeled $1-14$) involved in the definition of $V_p^{\{\mu_{p,q}\}}$ when flipping the middle plaquette starting from this particular initial configuration. Red rectangles correspond to the pair projector terms involved in $V_p^{\{\mu_{p,q}\}}$. Note that the spins of the involved Majorana modes are not shown in the figure.  }
\label{xp}
\end{figure}

Now we write the decoration part of the Hamiltonian as
\begin{align}
&H_{\text{decorate}} \nonumber \\
&= -\sum_{\substack{\langle \overrightarrow{vw} \rangle \\ t(v) = t(w)}} [ i D_{\langle \overrightarrow{vw} \rangle}W_{vw}^+\gamma_v^{\uparrow} \gamma_w^{\downarrow} 
+i D_{\langle \overrightarrow{vw} \rangle}W_{vw}^-\gamma_v^{\downarrow} \gamma_w^{\uparrow}
%
)] \nonumber \\
&\quad -\sum_{\substack{\langle \overrightarrow{vw} \rangle \\ t(v) \neq t(w)}} [ i D_{\langle \overrightarrow{vw} \rangle}\bigg( \frac{1+\tau_{f}^z}{2} \bigg) \gamma_v^\downarrow \gamma_w^\downarrow +  i D_{\langle \overrightarrow{vw} \rangle}\bigg( \frac{1-\tau_{f}^z}{2} \bigg) \gamma_v^\uparrow \gamma_w^\uparrow  \nonumber \\
& \qquad  \qquad  \qquad +i \bigg( \frac{1-D_{\langle \overrightarrow{vw} \rangle}}{2} \bigg) (\gamma_v^{\uparrow} \gamma_w^{\uparrow} +\gamma_v^{\downarrow} \gamma_w^{\downarrow})],
\end{align}
where $t(v)$ (resp. $t(w)$) denotes the triangular face that includes the vertex $v$ (resp. $w$). 
$H_{\text{tunnel}}$ can be defined by 
\begin{align}
H_{\text{tunnel}} = \sum_{p} \tau_p^x X_p,
\end{align}
where the sum over $p$ only involves the plaquettes, not the triangles. The plaquette term $X_p$ rearranges the Majorana chains to comply with the domain wall decoration rules defined above after $\tau_p^x$ is applied. Specifically, 
\begin{equation}
X_p = \sum_{\substack{ \mu_p=\pm 1\\ \{\mu_q=\pm 1\} }} V_p^{\{\mu_{p,q}\}}\Pi_p P_p^{\{\mu_{p,q}\}},
\end{equation}
where the sum over $\{\mu_q =\pm 1\}$ denotes the summation over all the adjacent plaquette spin configurations around $p$ . Note that by using the ``majority rule'', one can extend the spin configuration from plaquettes to triangles. 
The operators $P_p^{\{\mu_{p,q}\}}$ and $\Pi_p$ are projectors: $P_p^{\{\mu_{p,q}\}}$ projects onto bosonic spin states with precisely $\tau_p^z=\mu_p$ and $\tau_q^z=\mu_q$, and $\Pi_p$ projects onto states in the fermionic Hilbert space that conform to those spin configurations:
\begin{equation}
P_p^{\{\mu_{p, q}\}} =\bigg( \frac{1+\tau_p^z\mu_p}{2}\bigg) \prod_{\{q\}} \bigg( \frac{1+\tau_q^z\mu_q}{2} \bigg)
\end{equation}
\begin{align}
&\Pi_{p} =\prod_{\substack{\langle \overrightarrow{vw} \rangle \in \partial^{\prime} p \\ t(v) = t(w)}} D_{\langle \overrightarrow{vw} \rangle} \bigg[ W_{vw}^+ \bigg(\frac{1+i\gamma_v^{\uparrow} \gamma_w^{\downarrow}}{2} \bigg)+ W_{vw}^- \bigg(\frac{1+i\gamma_v^{\downarrow} \gamma_w^{\uparrow}}{2} \bigg) \bigg] \nonumber \\
& \prod_{\substack{\langle \overrightarrow{vw} \rangle \in \partial^{\prime} p \\ t(v) \neq t(w)}} \bigg
\{ \bigg(\frac{1-D_{\langle \overrightarrow{vw} \rangle}}{2} \bigg) \bigg(\frac{1+i\gamma_v^{\uparrow} \gamma_w^{\uparrow}}{2} \bigg) \bigg(\frac{1+i\gamma_v^{\downarrow} \gamma_w^{\downarrow}}{2} \bigg) +
\nonumber \\
& D_{\langle \overrightarrow{vw}  \rangle}\bigg[ \bigg(\frac{1+\tau_{f_{vw}}^z}{2} \bigg)\bigg(\frac{1+i\gamma_v^{\downarrow} \gamma_w^{\downarrow}}{2} \bigg) +\bigg(\frac{1-\tau_{f_{vw}}^z}{2} \bigg)\bigg(\frac{1+i\gamma_v^{\uparrow} \gamma_w^{\uparrow}}{2} \bigg)\bigg] \bigg\}\nonumber \\
\label{defproj}
\end{align}
Here $\partial^{\prime}p$ includes the 36 Majoranas in the triangles surrounding the plaquette $p$, as shown in Fig.\ref{xp}(a). The first line and third line of Eq.\eqref{defproj} enforce the pairing of Majorana modes on the domain wall, and the second line of Eq.\eqref{defproj} enforces the pairing of Majorana modes away from the domain wall.



The third part in the definition of $X_p$ is
\begin{align}
V_p^{\{\mu_{p,q}\}} = 2^{-\frac{n+1}{2}} &(1+is_{2,3}\gamma_2^{\sigma_2}\gamma_3^{\sigma_3})(1+is_{4,5}\gamma_4^{\sigma_4}\gamma_5^{\sigma_5})\dots \nonumber \\ &(1+is_{2n,1}\gamma_{2n}^{\sigma_{2n}}\gamma_1^{\sigma_1}). 
\label{Xpdw}
\end{align}
which takes the initial fermion configuration $|\Psi_i\rangle$ determined by $\Pi_p$ corresponding to a fixed bosonic configuration determined by $P_p^{\{\mu_{p,q}\}}$, and maps it to $|\Psi_f\rangle$. The constant in the front is chosen so that $\ket{\Psi_f}$ has the same norm as $\ket{\Psi_i}$. The labels $\sigma_i$ ($i=1,2...2n$) can take values $\uparrow$ and $\downarrow$, specifying the spins of the Majorana modes, and are determined by the bosonic spin configuration on and around the plaquette $p$ following the aforementioned decoration rules. The Majorana modes $\gamma_i$ are arranged so that the initial state satisfy $is_{2i-1,2i}\gamma_{2i-1}^{\sigma_{2i-1}} \gamma_{2i}^{\sigma_{2i}}=1$. Then $V_p^{\{\mu_{p,q}\}}$ maps this state into a state $|\Psi_f\rangle$ with $is_{2i,2i+1}\gamma_{2i}^{\sigma_{2i}} \gamma_{2i+1}^{\sigma_{2i+1}}=1$. Here $s_{i,j}=1$ if the edge $\langle v_iv_j \rangle$ points from $v_i$ to $v_j$ and $s_{i,j}=-1$ otherwise. A pictorial illustration is given in Fig.\ref{xp}(b).

$V_p^{\{\mu_{p,q}\}}$ defined above determines the relative weight and phase factor of different configurations. With repeated application of $V_p$ and $\tau^x_p$, we can start from any initial configuration (including both boson and fermion degrees of freedom) satisfying $H_{\text{decorate}}$, and reach any other final configuration. The total ground state wave function is then a superposition of all the configurations obtained in this way. The fact that the relative weight and phase factor of different configurations can be uniquely and consistently determined is guaranteed by the commutativity of different $V_p$ terms, which we prove in Appendix \ref{commute}. Moreover, as we discuss in Appendix \ref{HTR}, the Hamiltonian as defined is time reversal invariant and ensures the time reversal invariance of the ground state wave function.


\textit{Why $T^2=1$ fermion does not work} -- 
We now discuss why our decoration procedure discussed above does not work for spinless fermions with $T^2=1$. In particular, we will argue that if one decorates the time reversal domain walls with spinless Majorana chains, then the requirement of time reversal invariance for the wave function is not compatible with the requirement that any two decorated domain wall configurations in the superposition have the same fermion parity.

Let the spinless complex fermion on a short bond $l = \langle \overrightarrow{vv^{\prime}} \rangle$ be created and annihilated by operators $c_l$ and $c_l^{\dagger}$, respectively. We first represent the complex fermion by a pair of Majorana modes
$\gamma_{v} = c^{\dagger}_{l} + c_{l}$,
$\gamma_{v^{\prime}} = i(c^{\dagger}_{l}-c_{l})$
located at vertices $v$ and $v^{\prime}$, respectively. Under time reversal, $T: c_l \rightarrow c_l$.
Written in terms of the Majorana modes, we have:
\begin{align}
T: \gamma_v \rightarrow \gamma_v, \ \ \ \gamma_{v^{\prime}} \rightarrow -\gamma_{v^{\prime}}.
\end{align}

We may decorate the time-reversal domain walls with Majorana chains in a way similar to the $T^2=-1$ case. Away from the domain wall, we pair up Majorana modes that share a short bond $l = \langle \overrightarrow{vv^{\prime}} \rangle$ as $i\gamma_v \gamma_v^{\prime}$. 
On a domain wall, we pair up Majorana modes that share a long bond $\tilde{l} = \langle \overrightarrow{vw} \rangle$ as $i\gamma_v\gamma_w$. 

However, there is an issue with the above pairing rules, because it does not preserve time-reversal invariance. In particular, let us consider the pairing of Majorana modes that shares a long bond $\tilde{l} = \langle \overrightarrow{vw} \rangle$ on a domain wall. Denote by $v^{\prime}$ the vertex that shares a short bond with $v$, and $w^{\prime}$ the vertex that shares a short bond with $w$. For the specific Kasteleyn orientation we are working with, the short bonds $\langle vv^{\prime} \rangle$ and $\langle ww^{\prime}\rangle$ must have opposite Kasteleyn orientations. This implies that $\gamma_v$ and $\gamma_w$ transform identically under time reversal, which renders the coupling term $i\gamma_v \gamma_w$ odd under time reversal.

\begin{figure}
\centering
\includegraphics[width = \linewidth]{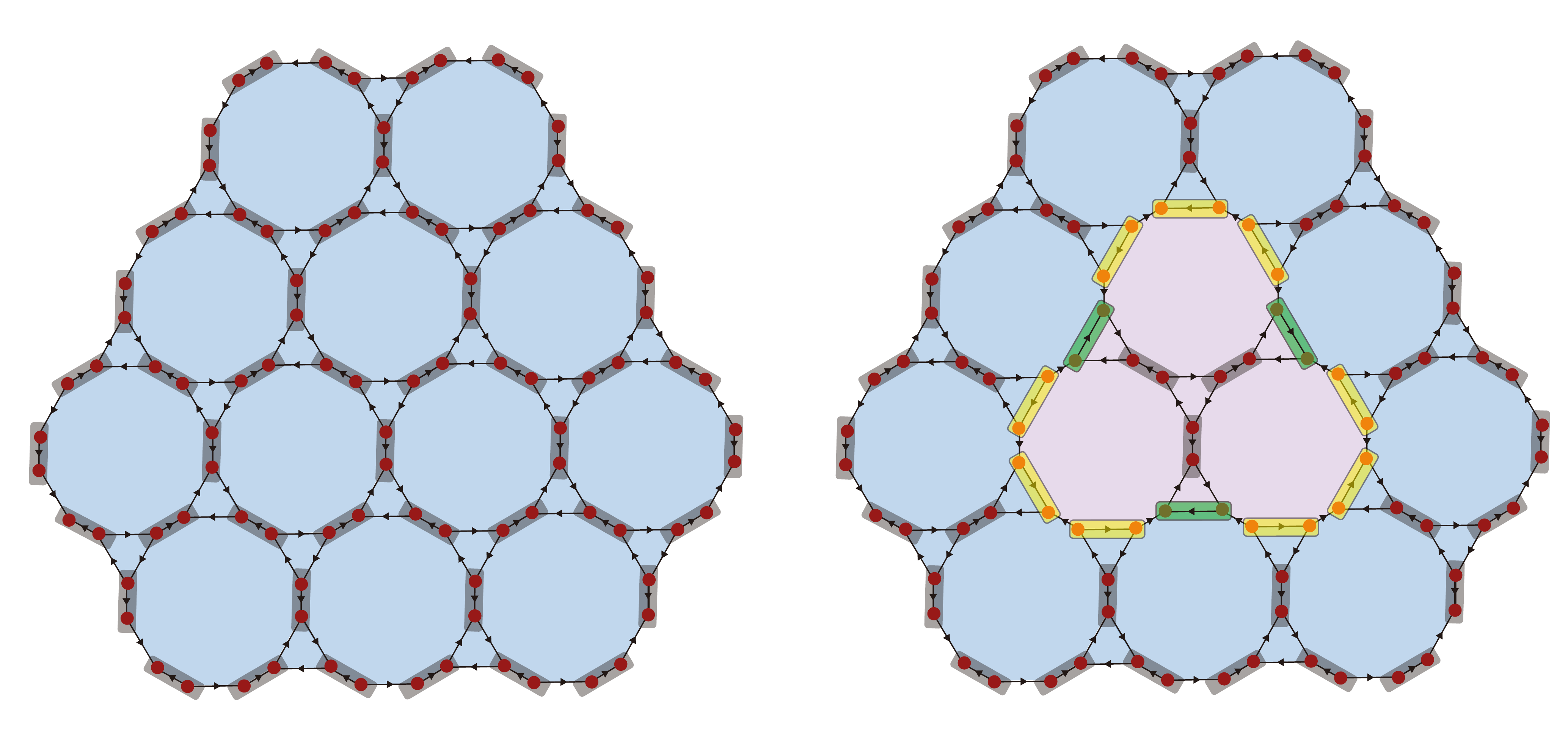}
\caption{Two configurations for spinless Majorana modes with opposite fermion parity. Extra minus signs are added to the coupling on the green bonds according to the modified coupling rule.}
\label{fparity}
\end{figure}

One may try to resolve this issue by adding a minus sign to the coupling when the left hand side of the long bond is in the $\ket{1}$ state. 
But this inevitably breaks the fermion parity invariance. Consider the two plaquette spin configurations in Fig.\ref{fparity}. Due to the Kasteleyn orientation, the two configurations will have the same fermion parity if we stick to the original coupling rule which breaks time-reversal invariance. The modified coupling rule introduces some extra minus signs into the fermion parity of the second configuration and the number of minus signs is exactly equal to the number of clockwise oriented bonds on the domain wall, which is three in this case. Therefore, with the modified coupling rules, the two configurations have opposite fermion parity.

We have thus argued that our decorated domain wall construction for $T^2=-1$ fermions cannot be generalized consistently to $T^2=1$ which incorporates both time-reversal invariance and fermion parity invariance. This is consistent with the fact that there are no nontrivial fermionic short-range entangled phases with $T^2=1$.


\textit{Conclusion} -- We have constructed an exactly solvable model for 2D topological superconductor with time reversal symmetry by decorating time reversal domain walls with spinful Majorana chains. Although our presentation focuses on the Honeycomb lattice, the construction actually works for any trivalent lattice using the same convention as defined in this paper. One can further ask whether it is possible to have a similar construction for 2D topological insulator and 3D topological insulator and superconductor. We leave these questions for future study. 


\textit{Acknowledgment} --
X.C. would like to thank Lukasz Fidkowski, Qing-Rui Wang, Zheng-Xin Liu and Jason Alicea for inspiring discussions and to the Kavli Institute for Theoretical Sciences for hosting when some of the discussion happened. X.C. is supported by National Science Foundation under
award number DMR-1654340, the Walter Burke Institute for Theoretical Physics and the Institute for Quantum Information and
Matter. S.-Q.N.  is supported by NSFC (Grant Nos.11574392), the Ministry of Science and Technology of China (Grant No. 2016YFA0300504), and the Fundamental Research Funds for the Central Universities and the Research Funds of Renmin University of China (No. 15XNLF19). X.C. and Z.W. would also like to thank the Institute for Advanced Study at Tsinghua University for hosting when this paper was being written.


\bibliography{ref}


\appendix

\section{The Hamiltonian terms commute with each other}
\label{commute}

The Hamiltonian in our model is a sum of commuting projectors. It is straightforward to see that all the $H_{\text{decorate}}$ terms commute and that all the $H_{\text{tunnel}}$ terms commute with all the $H_{\text{decorate}}$ terms. In this section, we prove that all the $H_{\text{tunnel}}$ terms commute with each other.

To prove that any pair of plaquette operators $\tau_{p_1}^x X_{p_1}$ and $\tau_{p_2}^x X_{p_2}$ commute,  it is equivalent to prove that for any state in the Hilbert space, the final states are the same independent of the order of the plaquette operator action. Namely,
\begin{align}
\tau_{p_1}^x X_{p_1}\tau_{p_2}^x X_{p_2}|\Psi\rangle=\tau_{p_2}^x X_{p_2}\tau_{p_1}^x X_{p_1}|\Psi\rangle.
\end{align}
For non-adjacent $p_1$ and $p_2$, these two terms involve different spins and Majoranas and act on the state independently, so they obviously commute. However, for adjacent $p_1$ and $p_2$, some of the Majorana modes that the two plaquette operators act on are the same, and it is not obvious whether they commute or not. Since $X_p$ by construction, guarantees that the Majorana configurations match the plaquette spin configurations, and the plaquette spin configuration is independent of the order in which we apply the plaquette operators, the final configuration of the Majorana modes are actually the same, but the fermionic state can differ by a complex phase, i.e., the plaquette operators commute up to a complex phase. As we will argue below, such complex phases are actually all equal to zero, and the plaquette operators commute exactly. 

Recall that $P_p^{\{\mu_{p,q}\}}$ projects onto the spin configuration of $\{ \mu_{p,q}\}$ and $\Pi_p$ projects onto the fermonic subspace that conforms to such spin configuration, so we only need to consider those states whose fermion parts  match the spin configurations. We denote such states as $|\Psi_{\{\mu_{p,q}\}}\rangle \otimes |\Psi_{spin}\rangle$. For adjacent $p_1$ and $p_2$, it is sufficient to consider states of the form $|\Psi_{\{\mu_{p,q}\}}\rangle \otimes |\tau_1, \tau_2, ..., \tau_N \rangle $. We compute 
\begin{align}
& \text{ }\tau_{p_1}^x  X_{p_1} | \Psi_{\{\mu_{p,q}\}}\rangle \otimes | \tau_1, \tau_2, ..., \tau_N \rangle  \nonumber \\
&= V_{p_1}^{\{\mu_{p,q}\}} | \Psi_{\{\mu_{p,q}\}}\rangle \otimes | \tau_1', \tau_2 , ..., \tau_N  \rangle \nonumber \\
&\propto | \Psi_{\{\mu^1_p, \mu_q\}} \rangle \otimes | \tau'_1, \tau_2 , ..., \tau_N  \rangle 
\end{align}

 $| \Psi_{\{\mu_{p,q}\}}\rangle$ and $| \Psi_{\{\mu^1_p, \mu_q\}}\rangle$ denote the same Majorana configurations except some, denoted by  
$\gamma_1^{\sigma_1}, \gamma_2^{\sigma_2},... \gamma_{2n}^{\sigma_{2n}}$, around the plaquette $p_1$. More explicitly, we arrange the Majorana modes so that $i s_{2i-1,2i} \gamma_{2i-1}^{\sigma_{2i-1}} \gamma_{2i}^{\sigma_{2i}}$$|\Psi_{\{\mu_{p,q}\}} \rangle = |\Psi_{\{\mu_{p,q}\}} \rangle$, while $i s_{2i,2i+1} \gamma_{2i}^{\sigma_{2i}} \gamma_{2i+1}^{\sigma_{2i+1}} |\Psi_{\{\mu^1_p, \mu_q\}} \rangle = |\Psi_{\{\mu^1_p, \mu_q\}} \rangle$. In this case, the operator $V_p^{\{\mu_{p,q}\}}$ is exactly of the form in Eq.\eqref{Xpdw}:
\begin{align}
V_{p_1}^{\{\mu_{{p_1},q}\}} = 2^{-\frac{n+1}{2}} &(1+is_{2,3}\gamma_2^{\sigma_2}\gamma_3^{\sigma_3})(1+is_{4,5}\gamma_4^{\sigma_4}\gamma_5^{\sigma_5})\dots \nonumber \\ &(1+is_{2n,1}\gamma_{2n}^{\sigma_{2n}}\gamma_1^{\sigma_1}).
\end{align}

Note that the choice of $\{ \sigma_i(\{\mu_{{p_1},q}\})\}$ depends on the plaquette spin configuration. 
This point becomes important when  considering two adjacent plaquettes. Now turn to two adjacent plaquttes $p_1$ and $p_2$ and we consider first flipping the $p_1$ spin and then the $p_2$ spin:
\begin{align}
&\tau_{p_2}^x X_{p_2}\tau_{p_1}^x X_{p_1}| \Psi_{\{\mu_{p,q}\}}\rangle \otimes | \tau_1, \tau_2, ... \rangle \nonumber \\
&= V_{p_2}^{\{\mu^1_{p,q}\}}V_{p_1}^{\{\mu_{p,q}\}} | \Psi_{\{\mu_{p,q}\}}\rangle \otimes | \tau_1', \tau_2', ... \rangle,  
\label{xp12}
\end{align} 
versus first acting on $p_2$ and then $p_1$:
\begin{align}
&\tau_{p_1}^x X_{p_1}\tau_{p_2}^x X_{p_2}| \Psi_{\{\mu_{p,q}\}}\rangle \otimes | \tau_1, \tau_2, ... \rangle \nonumber \\
&= V_{p_1}^{\{\mu^2_{p,q}\}}V_{p_2}^{\{\mu_{p,q}\}} | \Psi_{\{\mu_{p,q}\}}\rangle \otimes | \tau_1', \tau_2',... \rangle  
\label{xp21}
\end{align} 

\begin{figure}
\centering
\includegraphics[width = \linewidth]{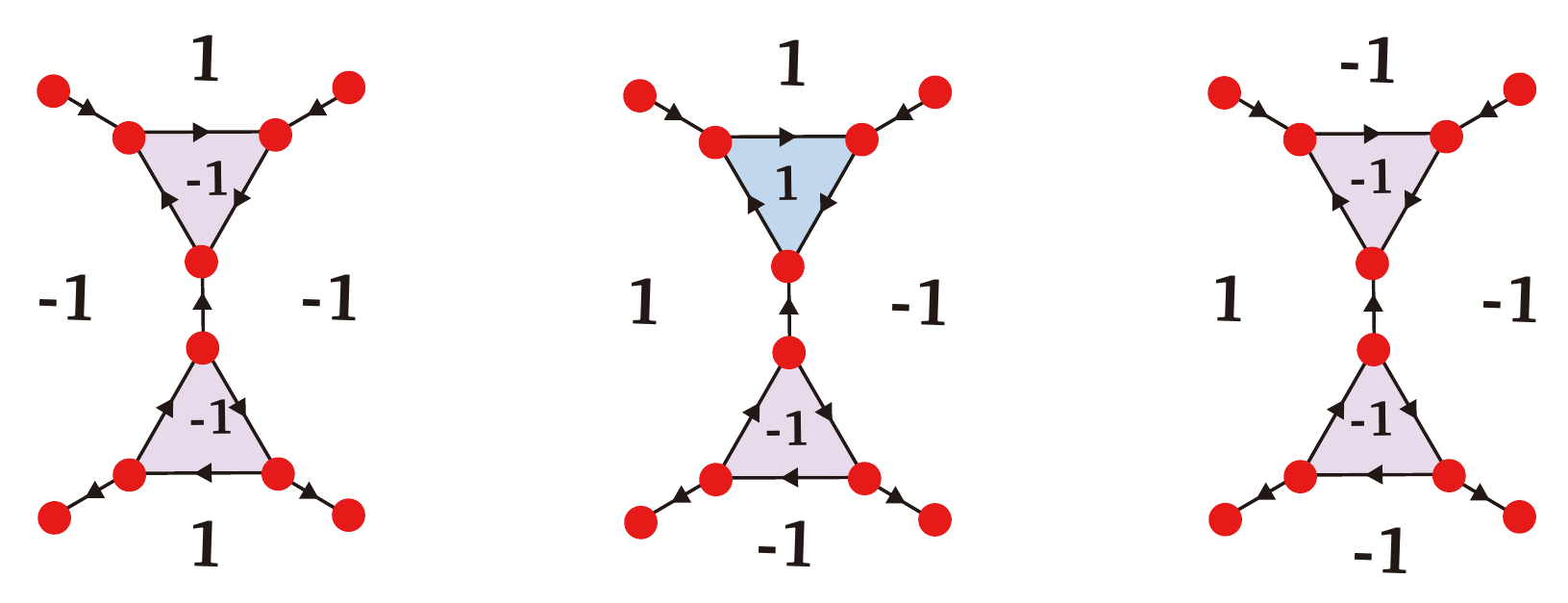}
\caption{The three relevant spin configurations when proving the commutativity of the plaquette operators.}
\label{three_cases}
\end{figure}

To prove that $\tau_{p_1}^x X_{p_1}$ and $\tau_{p_2}^x X_{p_2}$ commute is now equal to prove that the final states in (\ref{xp12}) and (\ref{xp21}) are exactly the same, not just the same up to a phase factor. To show this, we can use the following procedure. First, we notice the identity $P|\Psi_{\{\mu_{p,q}\}}\rangle=|\Psi_{\{\mu_{p,q}\}}\rangle$, where $P$ is the projector onto the fermionic state $|\Psi_{\{\mu_{p,q}\}}\rangle$. Using this identity, we can prove that $\tau_{p_1}^x X_{p_1}$ and $\tau_{p_2}^x X_{p_2}$ commute by simply proving that
\begin{align}
V_{p_1}^{\{\mu^2_{p,q}\}}V_{p_2}^{\{\mu_{p,q}\}}P=V_{p_2}^{\{\mu^1_{p,q}\}}V_{p_1}^{\{\mu_{p,q}\}}P,
\label{12=21}
\end{align}
where $\{\mu_{p,q}\}$ labels the initial spin configuration, and $\{\mu^{1}_{p, q}\}$ (resp. $\{\mu^{2}_{p, q}\}$) labels the spin configuration after the spin on plaquette $p_1$ (resp. $p_2$) is flipped. $p_1$ and $p_2$ share two triangles and one short bond, as seen in Fig.(\ref{three_cases}). In Eq.\eqref{12=21}, projectors which do not act on the Majorana modes on the two triangles commute obviously. Projectors that do act on the shared triangles may fail to commute. Since the configuration of the Majorana modes on the shared triangles depend on the spin configurations of $p_1$, $p_2$, and the two plaquettes bordering both $p_1$ and $p_2$, we may enumerate all the possible $2^4=16$ spin configurations on these $4$ plaquettes and explicitly check that Eq.\eqref{12=21} holds. We find that the $16$ cases essentially reduce to the three cases listed in Fig.\ref{three_cases} by symmetry arguments and similarity in proof techniques. A straightforward although lengthy calculation shows that Eq.\eqref{12=21} indeed holds for these three cases.

\section{Time Reversal Invariance of the Hamiltonian and the Wave Function}
\label{HTR}

Recall that we define the time reversal operation on the spins and fermions as $T=\prod \tau_x \otimes \prod (i\sigma_y) K$ with $K$ the complex conjugation operator. 

Under time reversal, both $D_{\overrightarrow{vw}}$ and $i(\gamma_v^{\uparrow} \gamma_w^{\uparrow} +\gamma_v^{\downarrow} \gamma_w^{\downarrow})$ are even. $W_{vw}^+i\gamma_v^\uparrow \gamma_w^{\downarrow}$ maps to $W_{vw}^-i\gamma_v^\downarrow \gamma_w^{\uparrow}$, and $( \frac{1+\tau_{f}^z}{2}) \gamma_v^\downarrow \gamma_w^\downarrow $ maps to $( \frac{1-\tau_{f}^z}{2} ) \gamma_v^\uparrow \gamma_w^\uparrow$. Therefore $H_{\text{decorate}}$ is time reversal invariant. It is not obvious that the tunnelling term is also time reversal invariant, we need to check it explicitly. First, the spin term $\tau_p^x$ is invariant under time reversal. Similar to $H_{\text{decorate}}$, it is obvious that the $\Pi_p$'s are even under time reversal. $P_p^{\{\mu_q, \mu_q\}}$ transforms to its time reversal counterpart because $T P_p^{\{\mu_q, \mu_q\}} T^{-1}= P_p^{\{-\mu_q, -\mu_q\}} $.  It can be explicitly checked that  $V_{p}^{\{\mu_{p,q}\}}$ also transforms into  its time reversal counterpart under time reversal.  Therefore, we see that
\begin{align}
TV_{p}^{\{\mu_{p,q}\}} \Pi_p P_{p}^{\{\mu_{p,q}\}} T^{-1}=V_{p}^{\{-\mu_p, -\mu_q\}} \Pi_p P_{p}^{\{-\mu_p, -\mu_q\}}.
\end{align}
 Although $X_{p}^{\{\mu_{p,q}\}} \Pi_p P_{p}^{\{\mu_{p,q}\}} $ alone is not time reversal invariant,  the sum  of all configurations of $\{ \mu_p, \mu_q\}$ is invariant under time reversal.
 
 Finally, let us come back to prove that the ground state wave function is time-reversal invariant. It suffices to prove that the weights of two configurations related by time reversal are complex conjugate of each other. Let us consider a fermionic state $\ket{\Psi_f}$ obtained by acting a sequence of plaquette operators on the initial fermionic state $\ket{\Psi_i}$ associated with the plaquette spin configuration where $\tau_p^z = 1$ for all $p$: $\ket{\Psi_f} = V_{p_1}V_{p_2}\dots V_{p_n}\ket{\Psi_i}$. The fermionic state $\ket{\Psi_f^T}$ associated with the time-reversal partner of this configuration can be obtained by acting another sequence of plaquette operators on the initial fermionic state: $\ket{\Psi_f^T} = V_{p_1^{\prime}}V_{p_2^{\prime}}\dots V_{p_m^{\prime}}\ket{\Psi_i}$, where $p_1^{\prime} \cup p_2^{\prime}\cup \dots \cup p_{m}^{\prime}$ form the complementary region of $p_1 \cup p_2 \cup \dots \cup p_n$. Note that the boundary of both regions agree. Using similar tricks as in Eq.(A11) of Ref.\onlinecite{Tarantino2016} for spinless fermions, we found that both $V_{p_1}V_{p_2}\dots V_{p_n}$ and $V_{p_1^{\prime}}V_{p_2^{\prime}}\dots V_{p_m^{\prime}}$ can be reduced to the product of a sequence of projectors which act only on the Majoranas lying on the boundary of the region $p_1 \cup p_2 \cup \dots \cup p_n$:
\begin{align}
V_{p_1}V_{p_2}\dots V_{p_n} = 2^{-\frac{n+1}{2}}& (1+is_{2,3}\gamma_2^{\sigma_2}\gamma_3^{\sigma_3})(1+is_{4,5}\gamma_4^{\sigma_4}\gamma_5^{\sigma_5})\nonumber \\
&\dots (1+is_{2n,1}\gamma_{2n}^{\sigma_{2n}}\gamma_1^{\sigma_1}), 
\label{boundaryproj}\\
V_{p_1^{\prime}}V_{p_2^{\prime}}\dots V_{p_m^{\prime}} = 2^{-\frac{n+1}{2}}& (1+is_{2,3}\gamma_2^{\bar{\sigma}_2}\gamma_3^{\bar{\sigma}_3})(1+is_{4,5}\gamma_4^{\bar{\sigma}_4}\gamma_5^{\bar{\sigma}_5})\nonumber \\
&\dots (1+is_{2n,1}\gamma_{2n}^{\bar{\sigma}_{2n}}\gamma_1^{\bar{\sigma}_1}). \label{boundaryproj2}
\end{align}
Furthermore, both $p_1 \cup p_2 \cup \dots \cup p_n$ and $p_1^{\prime} \cup p_2^{\prime}\cup \dots \cup p_{m}^{\prime}$ are in the $\tau_p^z = -1$ configuration. Therefore, by the coupling rules we introduced earlier, $\sigma_i$ and $\bar{\sigma}_i$ must be the opposite of each other for $i = 1,2,\dots,2n$. Hence Eq.\eqref{boundaryproj} and Eq.\eqref{boundaryproj2} can be mapped into each other term by term under time reversal. Hence the weights associated with $\ket{\Psi_f}$ and $\ket{\Psi_f^T}$ are complex conjugate of each other.


\end{document}